# Uncertainty and information in classical mechanics formulation. Common ground for thermodynamics and quantum mechanics

Adrián Faigón*

Device Physics Laboratory - Departamento de Física - Facultad de Ingeniería-
Universidad de Buenos Aires.

Mechanics can be founded on a principle relating the uncertainty δq in the trajectory of an observable particle to its motion relative to the observer. From this principle, p. δq =const., p being the q-conjugated momentum, mechanical laws are derived and the meaning of the Lagrangian and Hamiltonian functions are discussed. The connection between the presented principle and Hamilton's Least Action Principle is examined.

Wave mechanics and Schrödinger equation appear without additional assumptions by choosing the representation for δq in the case the motion is not trajectory describable. The Cramer-Rao inequality serves that purpose. For a particle hidden from direct observation, the position uncertainty determined by the enclosing boundaries leads to thermodynamics in a straightforward extension of the presented formalism.

The introduction of uncertainty in classical mechanics formulation enables the translation of mechanical laws into the wide ranging conceptual framework of information theory. The boundaries between classical mechanics, thermodynamics and quantum mechanics are defined in terms of informational changes associated with the system evolution. As a direct application of the proposed formulation upper bounds for the rate of information transfer are derived.

--------

*A. Faigón is research fellow at the CONICET (Consejo Nacional de Investigaciones Científicas y Técnológicas). afaigon@fi.uba.ar.

## I. INTRODUCTION

Mechanics is the basis of the physics building. In spite of their formal equivalence, each formulation of classical mechanics, Newton, Euler-Lagrange, Hamilton, Jacobi, contributes to clarify the fundamental concepts on which physics relies, and mediates the connection between its different branches [1,2].

Thermodynamics (TD) links to mechanics, following Gibbs and Boltzmann, through statistics and proper hypothesis on systems containing a large number of unobservable mechanical entities [3-5]. The relationship between classical mechanics (CM) and quantum mechanics (QM) is not that straightforward. From its very beginning, efforts have been made to interpret QM in a classical framework. This includes the works of Madelung, de Broglie, Bohm, Nelson, among others, with hydrodynamics, pilot wave, quantum potential, and stochastic mechanics developments [6-9]. A new attempt has been carried out in the last ten years through different ways of introducing the Fisher information into classical equations to obtain the Schrödinger equation [10-13].

The links between Physics and Information Theory have been explored with fruitful results in the fields of thermodynamics and quantum mechanics [14,15]. The assumed infinite precision in classical mechanics formulation prevented its treatment in a similar way.

This contribution is an attempt to show that classical mechanics can be formulated in terms of the uncertainty associated with the description of observables. In addition to providing a new scope on mechanical laws, this formalism could help in smoothing conceptual bridges between the main branches of Physics.

The core of the article is in sections II.A to II.D where the principle on which present formulation of Mechanics relies, the constancy of the pδq-product, is introduced. We start by focusing on a special variation of the trajectory, δq = ε/p, and we show it gives rise to a variational principle from which classical mechanics derives. Next we investigate the consequences of deriving and expressing the main physical laws in classical mechanics, thermodynamics, and quantum mechanics, in terms of a finite δq corresponding to a small but finite ε. We show that consistency is achieved by giving δq the meaning of trajectory uncertainty.

TD appears as a natural extension of the present formulation of CM. The relationship between entropy, Lagrangian, Hamiltonian and action is, therefore, explored --Section II.H-- opening the way to extend the application range of Information Theory to CM in Section II.I.

The introduction of uncertainty in its fundamental description level, raises the question about the limits of CM. This is addressed in section II.J where the boundaries between CM, QM and TD are neatly defined in informational terms.

A by-product of the present formulation is a contribution to recent attempts to find the informational content in the Lagrangian function, and to derive the Schrödinger equation using Fisher information. The Lagrangian function, the meaning of which is perhaps the less clear among the mechanical quantities, is shown to have a strong connection to information as suggested elsewhere [16]. Within the present framework, the first order variation of the Lagrangian corresponding to the uncertainty interval is closely related to the rate of information transfer. The limits of the rate of information transfer are, therefore, explored and compared with published results on the same subject obtained via different approaches [17,18].

## II. THEORY

### A. The special variation δq

Let $L(q, \dot{q}, t)$ be the Lagrangian function for a mechanical system, meaning that the generalized coordinates have a time dependence q(t) satisfying

$$\frac{d}{dt}\left(\frac{\partial L}{\partial \dot{q}}\right) - \frac{\partial L}{\partial q} = 0 \quad , \quad (1)$$

the Lagrange equation of motion.

A virtual variation of an hypothetical trajectory q(t) to (q+δq)(t) causes a variation of L

$$\delta L = \frac{\partial L}{\partial q}.\delta q + \frac{\partial L}{\partial \dot{q}}.\delta \dot{q} \quad . \quad (2)$$

This means that for each proposed q(t) there is a corresponding variation δq(t) satisfying

$$\frac{\delta q}{\delta \dot{q}} = -\frac{\partial L/\partial \dot{q}}{\partial L/\partial q} \quad , \quad (3)$$

which nulls the first variation of L for all t [19]. For the actual trajectory, which satisfies eq. (1),

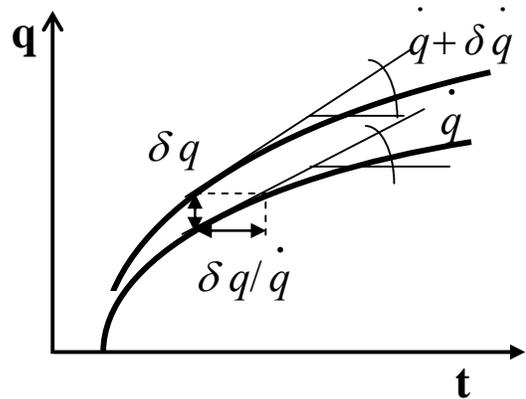

**Fig. 1.** The actual q(t) and varied (q+ δq)(t) trajectories

the denominator in the r.h.s. of Eq. (3) is the time derivative of the numerator, and the condition for zero δL can be rewritten

$$\frac{d}{dt}\left(\frac{\partial L}{\partial \dot{q}}.\delta q\right) = 0 \quad , \quad (4)$$

where

$$.\delta \dot{q} = \frac{d(q+\delta q)}{dt} - \frac{dq}{dt} = \frac{d(\delta q)}{dt} = (\dot{\delta q}) , \quad (5)$$

was used --Fig. 1--. I.e., the variation δq(t) that nulls δL when q(t) is the actual trajectory satisfies

$$\delta q = \varepsilon / p , \qquad (6)$$

where ε is an infinitesimal parameter with proper units. Last paragraph leads to the following principle.

### B. The differential form

A prescription to find the actual trajectory is:

Given a Lagrangian $L(q, \dot{q}, t)$, and initial conditions for q and $\dot{q}$,
Propose a trajectory q(t)
Obtain the momentum $p = \partial L / \partial \dot{q}$
Vary the trajectory in $\delta q(t) = \varepsilon / p$
Calculate the corresponding first order δL
Is zero? Then q(t) is the actual trajectory.

Proof:
Using Eqs. (2) and (5) the calculated δL is

$$\delta L = \frac{\partial L}{\partial q} \delta q + \frac{\partial L}{\partial \dot{q}} . (\dot{\delta q}) \qquad (7)$$

which is zero if

$$\frac{\partial L}{\partial q} = \frac{d}{dt} \left( \frac{\partial L}{\partial \dot{q}} \right) \qquad (8)$$

for, in this case, δL takes the form of the l.h.s. of eq. (4) which is zero for the special variation (6) used. Eq. (8) is Lagrange equation of motion. All the above is summarized in:

*The actual trajectory minimizes (makes stationary) L at each time instant with respect to variations δq(t) satisfying Eq. (6).*

### C. The integral form and the Principle of Least Action

The variation of the Hamiltonian action between the instants t1 and t2,

$$A \Big|_{t1}^{t2} = \int_{t1}^{t2} L \, dt \qquad (9)$$

is, after integrating by parts the term containing $\delta \dot{q}$, [1]

$$\delta A \Big|_{t1}^{t2} = p.\delta q \Big|_{t1}^{t2} + \int_{t1}^{t2} \left( \frac{\partial L}{\partial q} - \frac{d}{dt} \left( \frac{\partial L}{\partial \dot{q}} \right) \right).\delta q.dt \qquad (10)$$

Hamilton Principle of Least Action (PLA), stems from Eq. (10): The actual trajectory q(t), or equivalently, the trajectory which satisfies Lagrange equation, makes the action variation zero, for all variations of path δq(t) satisfying δq(t1) = δq(t2) = 0.
But in fact, it is sufficient to ask for a less tighten constraint, i.e. pδq(t1) = pδq(t2). Therefore the special variation (6) ensures δA=0 calculated between any two instants --there are no particular instants at which δq must be zero -- if q(t) is the real trajectory. Or:

*The actual trajectory minimizes (makes stationary) the action among the varied trajectories by variations of the form (6) between two any instants of time.*

Alternatively, as the actual trajectory zeroes the integral in Eq. (10) independently of the variation δq,

*The actual (classical) trajectory passing through (t1,q1) and (t2,q2) minimizes the action for all variations δq(t) provided pδq(t1) = pδq(t2).*

In this form, without conditions on the variation but at the ends t1 and t2, the principle is comparable to Hamilton's PLA, which appears to be the special case pδq(t1) = pδq(t2) = 0.

### D. The physical form

Up to this point δq is what in the texts is referred to as a mathematical variation. We will investigate in what follows if a physical meaning can be assigned to a finite δq(t) which tentatively will be called uncertainty in the trajectory.
This is suggested among others, by the principles stated above, meaning that at first order the trajectories varied according to (6) are indistinguishable --same value of L at given time, same action between any pair of time instants-- from the actual trajectory. Thus, let us assume there exists a small but finite value for ε, let call it f, such that the closest distinguishable trajectories are separated by δq satisfying

$$p \, \delta q = f = \text{const} \quad . \quad (11)$$

Or, better, as the trajectory concept becomes in this case blurred: the motion occurs within a δq satisfying eq. (11) centered in the function q(t) satisfying the Lagrange equation.

As each q(t) has an associated δq(t), we could make a conceptual inversion and put δq in the basis of the building as it contains all the information needed, apart from initial conditions, to determine the motion. For if we know δq as a function of position and/or time, p(t) is determined by Eq. (11) and the motion is solved.

Of course it makes sense if δq can be given any reality, or if some usefulness can be found for this entity. In what follows we will investigate the formal connection of eq. (11), which I will call the constancy of the p-δq product (CPDQ), with the classical formulation of classical mechanics, with thermodynamics and quantum mechanics.

### E. Classical mechanics

The classical formulation of classical mechanics is obtained from CPDQ by assuming that in spite of the uncertainty in position, the motion does occur along a trajectory, a continuous curve assigning a position q to each time instant t. In this case, the time derivative of eq. (11),

$$\dot{f} = \dot{p} \cdot \delta q + p \cdot (\dot{\delta q}) \quad , \quad (12)$$

can be written, using Eq. (5), as

$$\dot{f} = \dot{p} \cdot \delta q + p \cdot \delta \dot{q} \quad , \quad (13)$$

which compared with Eq. (2), and using the differential form of the principle, stating that on the actual trajectory df/dt = δL = 0, results in

$$p = \partial L / \partial \dot{q} \quad \text{and} \quad \dot{p} = \partial L / \partial q \quad , \quad (14a,b)$$

equivalent to Eq. (1), the Lagrangian equation of motion for the actual trajectory in the classical formulation.

### F. Mechanical quantities and Hamiltonian description

In view of the trajectory indeterminacy, the time derivative of any given quantity y, will be evaluated by

$$\dot{y} = \frac{\dot{q}}{\delta q} \cdot \bar{\delta} y \quad , \quad (15)$$

where $\bar{\delta} y$ is defined as the change in y while the particle moves from q to q+δq -the bar means here along the motion--, i.e. as it crosses the uncertainty interval, and $\delta q / \dot{q}$ the time interval along which this change occurs --Fig.1--. Applied to q itself, last definition implies [20]

$$\bar{\delta} q = \delta q \quad . \quad (16)$$

Replacing (16) in the first addend in (12), and (15) in the second one, one obtains [21]

$$\dot{p} \cdot dq = -\frac{p \dot{q}}{\delta q} \cdot d \delta q \quad , \quad (17)$$

which is the differential work W. From it,

$$\dot{p} = -p \dot{q} \frac{d \ln \delta q}{dq} \quad , \quad (18)$$

Newton's second law, wherein the r.h.s. stands for the force (generalized force); and, in case the forces are conservative ones, the potential energy is

$$dV \equiv -dW = p \dot{q} \cdot d \ln \delta q \quad . \quad (19)$$

Replacing now eq. (15) in both terms of Eq. (12), yields

$$\dot{f} = \dot{q} \cdot \bar{\delta} p + \frac{p \dot{q}}{\delta q} \cdot \bar{\delta} \delta q \quad (20)$$

The first term is the kinetic energy

$$dT \equiv \dot{q} \cdot dp \quad , \quad (21)$$

thus, equating the r.h.s. of expression (20) to zero --implied by CPDQ-- stands for the work-energy theorem.

Eq. (20) can be transformed --expressing the changes in δq in terms of changes in position q-- in

$$\dot{f} = \dot{q} \cdot \overline{\delta p} + p\dot{q} \cdot \frac{d \ln \delta q}{dq} \overline{\delta q} = \overline{\delta H} \quad , \quad (22)$$

which, in the conservative case --eq. (19)--, results in the change (along the motion) of a function H(q,p), the Hamiltonian function, H=T+V. From (20) and (16), the Hamiltonian equations of motion follow:

$$\dot{q} = \frac{\partial H}{\partial p} \quad \text{and} \quad \dot{p} = -\frac{\partial H}{\partial q} \quad , \quad (23 \text{ a,b})$$

completing the derivation of the classical formalism from the CPDQ. Equation (22) together with df/dt=0 is equivalent to the conservation of mechanical energy E, sum of the kinetic energy T and the potential energy V.

### G. A gate to Thermodynamics

Thermodynamics deals with systems of particles enclosed in a volume with no direct interaction with the surroundings but through the boundaries of the enclosure. The "mechanics" of the enclosed particle is as above but with δq determined by the linear dimension of the enclosure, of the order of and proportional to $Vol^{1/3}$ (or $(Vol/3N)^{1/3}$ for N particles).

We show in the following that with appropriate (natural) identifications of thermodynamic quantities with mechanical ones, one obtains a sort of "heat theorem" in Boltzmann's sense of proving [22]

$$dE = \theta \cdot dS - PdVol \quad , \quad (24)$$

where P is pressure, θ the absolute temperature, S entropy; on mechanical grounds. Or, in other terms, that there exists an integrating factor (1/θ) for the expresion dE+PdVol to become an exact differential (dS).

The (natural) identifications are --the simplest case of non-interacting particles is considered here--

$$dE \equiv \dot{q} \cdot dp \quad , \quad (25a)$$

$$\theta \equiv p\dot{q}/k \quad , \quad (25b)$$

$$P \equiv p\dot{q}/Vol \quad , \quad (25c)$$

or

$$PdVol \equiv p\dot{q}d\ln \delta q \quad , \quad (25d)$$

and

$$dS \equiv k.d \ln f \quad , \quad (26)$$

where k is Boltzmann constant and extensive quantities (differentials on the l.h.s.) are per degree of freedom; which replaced in Eq. (24) yield

$$d \ln f = d \ln p + d \ln \delta q$$

a differential form of eq. (11) proving the equivalence.[23]

### H. Extended Lagrangian and Hamiltonian. Action and entropy

As shown in sections II.A to II.D mechanics is characterized by the fact that the trajectory uncertainty, δq(t) that makes zero δL, is given by f=p.δq=const. In thermodynamics, trajectory uncertainty is imposed by the system boundaries, being therefore independent of p, freeing the value of both df/dt and δL which cease to be necessarily zero. The explicit relationship between both quantities, and consequently between Lagrangian and entropy is analyzed here.

Starting from the general variation of the mechanical Lagrangian function L(q,dq/dt), and remembering δ(dq/dt)= d(δq)/dt, which in the thermodynamic case means the rate of expansion, one gets

$$\delta L = \frac{\partial L}{\partial q}\delta q + \frac{\partial L}{\partial \dot{q}}(\dot{\delta q}) =$$
$$= \frac{\partial L}{\partial q}\delta q + \frac{d}{dt}(\frac{\partial L}{\partial \dot{q}}\delta q) - \frac{d}{dt}\frac{\partial L}{\partial \dot{q}}.\delta q \quad (27)$$

And, identifying $\partial L/\partial(dq/dt)$ with the conjugate momentum p,

$$\delta L = \dot{f} + (\frac{\partial L}{\partial q} - \dot{p})\delta q. \qquad (28)$$

Or, dividing by f and using (26),

$$\frac{1}{p}(\frac{\delta L}{\delta q} + \dot{p} - \frac{\partial L}{\partial q}) = \frac{\dot{S}}{k}, \qquad (29)$$

which relates the variational derivative of the Lagrangian function with the entropy rate of growth.

We show in what follows a simplification of the relationship between basic mechanical and thermodynamic quantities based on a direct extension of the Lagrangian defined in the form (13)

$$\delta L^e \equiv \dot{p}\,\delta q + p\,\delta \dot{q} \qquad . \qquad (30)$$

This is a non exact variation, as is the variation $\delta L$ of the mechanical Lagrangian function, but holds with the latter the same relationship existing between work and potential. The difference between them is

$$\delta L^e - \delta L = \left(\dot{p} - \frac{\partial L}{\partial q}\right)\delta q = \frac{1}{\dot{q}}\frac{\partial E}{\partial t}\delta q, \quad (31)$$

i.e. the virtual work of non-conservative agents of change of p, namely non-conservative forces and heat. Both quantities coincide therefore in conservative mechanics, having the extended Lagrangian $L^e$ a very simple meaning in thermodynamic evolutions,

$$\delta L^e = \frac{f}{k}.\dot{S} \qquad , \qquad (32)$$

as results from (29) and (31).

The Hamiltonian action is naturally extended by definition: multiplying both sides of eq. (32) times dt one obtains

$$\frac{d\delta A^e}{f} = \frac{dS}{k} \qquad , \qquad (33)$$

which expresses the relationship between action and entropy: the uncertainty variation of the action changes, in f units, as does the entropy in k units [24]. As does the Lagrangian, the extended action reduces to the Hamiltonian action in the mechanical case.

Completing the extension of main mechanical quantities, the Hamiltonian extension $H^e$ takes the same form defined in eq. (22) which reduces to $H^e$=T+V in the mechanical conservative case as shown, and satisfies with the extended $\delta L^e$ the known relationship between Lagrangian and Hamiltonian mechanical functions,

$$dH^e = d\left(\dot{q}\,p - L^e\right) \quad , \qquad (34)$$

taking into account eq. (17) defining the differential work. In his extended form, $dH^e$ coincides --eqs. (24) to (26)-- with $\theta dS$, i.e. with what is called reversible heat. From eqs. (13) for the extended Lagrangian and (20) for the extended Hamiltonian, the last relationship between both quantities is completed with

$$\bar{\delta}H^e = \delta L^e = \dot{f} \qquad , \qquad (35)$$

relating the variation of the extended Lagrangian corresponding to the trajectory variation $\delta q$ with the change in the extended Hamiltonian over the same interval along the trajectory $\bar{\delta}q$, and both with the entropy rate of change through (32).

**I. Mechanics and information**

Information theory entered Physics through the concept of entropy [25, 14, 15]. It does not permeate into Mechanics because by its very definition, information involves the change of some interval of uncertainty, this being a non-existent entity in the classic formulation which assumes infinite precision in the knowledge of every quantity. It naturally follows that in the present formulation, which relies, on the contrary, on the indeterminacy in the classical trajectory, an attempt be made to interpret Mechanics in terms of information.

Consistency is achieved by --naturally-- taking [14,15]

$$dI = -dS/k \qquad , \qquad (40)$$

and, in virtue of eqs. (26) and (11), the information can be decomposed in position and momentum information

$$dI \equiv dI_q + dI_p \quad ,$$

with

$$dI_q = \frac{dW}{k\theta} \quad \text{and} \quad dI_p = \frac{-dT}{k\theta} \quad , \quad \text{(41 a,b)}$$

where Eqs. (19), (21) and (25b) were used.

Consequently, conservation of mechanical energy, or, more general, the work-energy theorem which in turn derives from CPDQ, Eq. (11), mean that, in mechanical interactions, the mechanical information of each observable coordinate is conserved,

$$dI = 0 \quad , \quad (42)$$

suggesting the name Principle of Zero Information Exchange (PZIE) for CPDQ, Eq. (11), in this framework.

In informational terms, mechanical laws have very simple forms. It has been shown that kinetic and potential energies represent information on momentum and position respectively, multiplied by $k\theta$ [26], i.e., the informational representation of the Hamiltonian is

$$dH^e \equiv -k\theta \, (dI_q + dI_p) \quad . \quad (43)$$

Force is, therefore, $k\theta$ times the gradient of position information, and the conservation of mechanical energy is a direct expression of PZIE. So is the PLA as shown in the following sequence

$$0 = \int_{t1}^{t2} \dot{I} \cdot f \, dt = \int_{t1}^{t2} \frac{f}{k} \dot{S} \, dt = \int_{t1}^{t2} \delta L \, dt = \delta A \quad (44)$$

where first equality follows from Eq. (42), the second from Eq. (40), the third from Eq. (32), and $\delta L^e = \delta L$ corresponding to the mechanical case was used.

### J. The classical mechanics boundaries. Equilibrium, reversibility, trajectory condition.

Regarding the limits of validity of classical mechanics, it has been shown that it is the fixed relationship between δq and p prescribed by CPDQ what distinguishes it from thermodynamics.

There is, however, another restriction, namely the assumption stated in Eq. (16), $\bar{\delta} q = \delta q$, required to arrive at Newton's law, Eq. (18), for classical mechanical laws to hold.

Eq. (16) naturally follows from the definition of $\bar{\delta}$ as the change in the variable to which it is applied when the particle crosses the uncertainty interval, or passes from q to q+δq; hence, the change in position $\bar{\delta}$q is δq. But, strictly interpreted, this means δq(q) itself does not change along the interval, or, by successive applications, it is constant along the trajectory, and so is p, leading to the absence of dynamical changes. Dynamics requires lessening the constraint to allow gradual changes of δq, i.e.

$$\bar{\delta}\delta q << \delta q \quad , \quad (45)$$

which still permits to assign single values δq and p to the interval. Eq. (45) leads, by using (41), (19) and (21), to the equivalents:

$$\left|\bar{\delta} I_q\right| = \left|\frac{\bar{\delta} V}{k\theta}\right| << 1 \text{ or } \left|\bar{\delta} I_p\right| = \left|\frac{\bar{\delta} T}{k\theta}\right| << 1 . \text{ (46 a,b)}$$

Eq. (46) is a synthetic expression for the condition of applicability of the WKB approximation; or, in a broader sense, the condition for which, according to Ehrnfest's theorem, quantum expectation values of mechanical quantities behave like their classical analogues: The potential does not change too much along one wavelength, usually written λ(dV/dx)/2(E-V) << 1. This condition and its consequences are usually referred to as the Classical limit of Quantum Theory [27, 28]. Eq. (46a) may be rewritten

$$\left|\bar{\delta} I_q\right| = \left|\frac{\bar{\delta} V}{k\theta}\right| = \frac{p \cdot \dot{q}}{(k\theta)^2} \bar{\delta} V = \\ = \frac{f}{(k\theta)^2} \frac{dV}{dt} << 1 \quad ,(47)$$

where (15) and (25b) where used. The last inequality have the same form of Bohm's condition about how slow the change of the potential must be in order for a quantum system to evolve adiabatically (meaning continuously, without emission or absorption, in his context) $h/(\Delta Eo)^2 \cdot dV/dt \ll 1$ [29], $\Delta Eo$ the energy distance to processes, or trajectory describable processes.

If the condition is violated, no single value $\delta q$ describes the uncertainty along the interval $(q, q+\delta q)$ and classical mechanics fails. Quantum mechanics solves this difficulty by abandoning the trajectory representation while

|  | $\overline{\delta I_q}$ and $\overline{\delta I_p} \ll 1$ | ELSE |
|---|---|---|
| $\overline{\delta I} = \overline{\delta I_q} + \overline{\delta I_p} = 0$ | Classical Mechanics<br><br>Adiabatic Equilibrium Thermodynamics | Quantum Mechanics<br><br>(without classical approximation) |
| ELSE | Non-adiabatic Equilibrium Thermodynamics | Non-adiabatic Non-equilibrium (Classical or Quantum) Thermodynamics |

**Table I. Boundaries of the main description frameworks of Physics according to the information changes associated to the system evolution.**

the neighboring state. Both cases put a limit to the rate of change of the potential in the form

$$\frac{dV}{dt} \ll \frac{\Delta E}{\tau} \quad , \qquad (48)$$

being $\Delta E$ the relevant energy with which changes of potential are to be compared: (twice) the kinetic energy of the particle in the former, the quantum of radiating energy in the latter; and $1/\tau$ its associated frequency: $1/\delta q \cdot dq/dt = k\theta/f$ and $\Delta Eo/h$ respectively. The related thermodynamic problem of how slow a container wall has to move in order for the enclosed gas retain the equilibrium behavior shares the same principle, eq. (45), and derived equations. Using in (48) $dE/dt$ instead of $dV/dt$ as would have resulted if Eq. (46b) instead of (46a) had been used in the derivation, and rewriting it in terms of L the linear dimension of the enclosure, $L \approx N^{1/3} \cdot \delta q = N^{1/3} \cdot f/k\theta \cdot dq/dt$, i.e.: $dE/dt = dE/dL \cdot dL/dt$, $\Delta E \approx E \cdot \delta q/L = 3/2 \cdot N^{2/3} \cdot k\theta = 3/2 \cdot N \cdot f/L \cdot dq/dt$, and $\tau = f/k\theta$ as before, one gets the expected relationship $dL/dt \ll dq/dt$. Brillouin has derived from Bohm's relationship the more general result $dL/dt \ll c$ the velocity of a perturbation propagation in the medium, which is $dq/dt$ for a gas of non-interacting particles.[30]

What was shown in the preceding paragraph is that condition (45) or equivalents for the classical description to be valid, is the general condition for an evolution of a classical, quantum or thermodynamic system to be describable as a continuous succession of equilibrium states, i.e. quasi-static or reversible preserving CPDQ as shown in next section.
The above discussion is summarized in Table I which clarifies the boundaries of the fundamental branches of Physics in terms of informational changes [31].

**K. A gate to quantum mechanics**

In previous sections it has been shown that mechanics can be derived from
the CPDQ principle relating momentum with trajectory uncertainty. Classical mechanics laws are obtained if in spite of this uncertainty one attempts to express CPDQ in terms of varied trajectories of a point mass. In last section (J) we have shown the conditions for such a description to represent actual motions.

Quantum mechanics wave equations are the expressions of the same (now) general law, CPDQ, but instead of representing $\delta q$ by the width of the bundle of indistinguishable trajectories, it is represented by the "width" of a probability distribution of the particle position; a way of describing position uncertainty when the trajectory representation does not apply. The Cramer-Rao (CR) bound [32]

$$(\delta q)^2 \int \frac{P'^2}{P} dq \geq 1 \quad , \qquad (49)$$

where P is a probability density function of the observable q, and $(\delta q)^2$ is the variance estimation, provides a good starting point for finding such a distribution. We begin by asking

P and δq to satisfy the lowest CR bound. i.e. the equality in expression (49), which, in Cartesian coordinates, is written

$$\int \frac{P'^2}{P} dx' = \frac{1}{(\delta x)^2} = \frac{p^2}{f^2} = \frac{2m}{f^2} \cdot T$$
$$= \frac{2m}{f^2} \cdot (E - V(x))$$
(50)

where CPDQ, Eq. (11), and conservation of energy were used. The left hand side is the Fisher information associated with translations of the observable x with corresponding pdf P(x). δx is called in this context Fisher length [10,12]. Without abandoning the classical case, i.e. knowing the value of the kinetic energy at each position, it can be expressed

$$T(x) = E - V(x) =$$
$$= \int [E - V(x)] P(x') dx'$$
(51)

provided P is normalized to one, and Eq. (50) reads

$$0 = \int P \cdot [\frac{f^2}{2m} \frac{P'^2}{P^2} + V(x) - E] dx' .$$
(52)

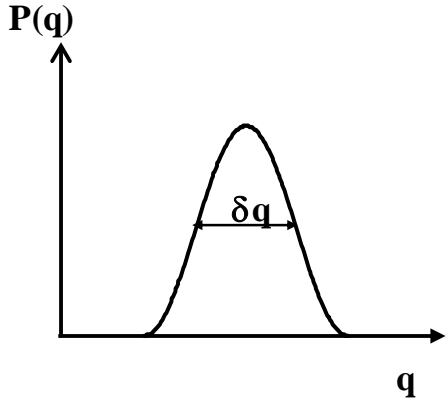

**Fig. 2. A function P(q) representing δq**

In addition, Cramer-Rao bound requires that zero is a minimum for the r.h.s. against variations of P, or

$$0 = \delta \int \frac{f^2}{2m} \frac{P'^2}{P} + P \cdot [V(x) - E] dx' .$$
(53)

where δ is here the usual sign for mathematical variation. Before writing down the Euler-Lagrange equation corresponding to eq. (53), let us generalize Cramer-Rao expression through the replacement of $P'^2/P$ by $4\psi'^*\psi'$ in Eq. (49) and the following ones, being ψ a complex amplitude satisfying $\psi^*\psi = P$. I.e. let us ask

$$(\delta x)^2 \int 4\psi'^*\psi' dx \geq 1$$
(54)

which reduces to the original Cramer-Rao's form in case ψ is real and yields the lowest value for δx in case it is not.[33] Eq. (53) is thus rewritten

$$0 = \delta \int \frac{2f^2}{m} \psi'^*\psi' + \psi^*\psi \cdot [V(x) - E] dx',$$
(55)

and the corresponding Euler-Lagrange equation for $\psi^*$ is

$$-\frac{2f^2}{m} \psi''(x') + \psi(x') \cdot V(x) = E\psi(x')$$
(56)

with solutions of the form

$$\psi(x, x') = e^{ik(x)x'}$$
(57a)

$$k(x) = [(E - V(x))/4C]^{\frac{1}{2}} .$$
(57b)

where $C = f^2/2m$. The above result should be read: Provided we know the kinetic energy at point x --meaning a definite p or δx--, the particle has at this position (or better in the interval δx centered at this position) an associated function of x' representing adequately δx.

If T(x) is not defined because V(x) changes strongly within δx --see Section J above-- then

$$T = \int [E - V(x')] P(x') dx'$$
(58)

which inserted in (51) and following the same procedure as above, yields

$$-\frac{2f^2}{m}\psi''(x') + V(x')\psi(x') = E.\psi(x'), \quad (59)$$

differing from Eq. (56) in that Eq. (59) is in the unique variable x'. Eq. (59) is Schrödinger time independent wave equation if $f=\hbar/2$ [34]. This is therefore the value of the constant in Eq. (11).

The relativistic, and time dependent SE follow immediately. First the relativistic version of SE is obtained by repeating the same arguments in Minkovsky space: Eq. (11) becomes

$$\delta \mathbf{s} \cdot \mathbf{I} = f \quad (60)$$

where

$$\delta \mathbf{s} \equiv (i\delta x, c\delta t) \text{ and } \mathbf{I} \equiv (ip, \frac{E}{c}) \quad (61a,b)$$

are the 4-vector (simplified here to one spatial dimension) for the worldline uncertainty, and momentum respectively. And Eq. (49) with the generalization of Cramer-Rao is

$$(\delta s)^2 \int 4\psi'^*\psi' dxdt \geq 1, \quad (62a)$$

with

$$\psi' = \partial\psi \equiv (i\frac{\partial}{\partial x} + \frac{\partial}{c\partial t})\psi. \quad (62b)$$

The condition on $\psi$ from Eqs. (62) and (60) is (equivalent to Eq. (55) above),

$$0 = \delta \int \psi'^*\psi' - \psi^*\psi \cdot \frac{m_o^2 c^2}{4f^2} dxdt, \quad (63)$$

where the relativistic energy momentum relationship, or $|\mathbf{I}|^2 = m_o^2 c^2$ was used. The Euler-Lagrange equation for $\psi^*$ is

$$0 = \psi'' + \frac{m_o^2 c^2}{4f^2}\psi$$
$$or \quad 0 = -\frac{\partial^2\psi}{\partial x^2} + \frac{\partial^2\psi}{c^2\partial t^2} + \frac{m_o^2 c^2}{4f^2}\psi \quad (64)$$

which is Klein-Gordon equation for $f=\hbar/2$.

Replacing in Klein Gordon equation $\psi=\psi_o.\exp(-im_oc^2t/\hbar)$ and taking the non-relativistic limit --equivalent to substracting out the rest energy and assuming that the remaining energies are small in comparison with it [35]--, yields

$$i\hbar\frac{\partial\psi_o}{\partial t} = -\frac{\hbar^2}{2m}\frac{\partial^2\psi_o}{\partial x^2}, \quad (65)$$

the time dependent SE for the free particle.

**L. The limiting rate of information transfer**

In a preceding section (II.J), the question of how slow the information change must occur for the system to remain in equilibrium was addressed. The opposite limit is considered here: how fast the information can be transferred. This is a question of actual interest, and enables direct comparison of results of the present formulation with known developments on this subject.

Eqs. (40), (32) and (35), lead to the following expressions for the rate of information transfer

$$\dot{I} = \frac{\delta L^e}{f.\ln 2}, \text{ and } \dot{I} = \frac{\overline{\delta}H^e}{f.\ln 2} \quad (66 \text{ a,b})$$

--the ln2 is to express the result in binary digits (bits)/sec--. The first expresses the rate of information transfer as the instant value of the extended Lagrangian uncertainty variation measured in f.ln2 units. The second allows to set the upper bound for the rate of information "emitted" by a system, or delivered by a signal of energy E, to

$$\dot{I} < \frac{E}{f.\ln 2} < \frac{4\pi E}{h.\ln 2}. \quad (67)$$

Last value, which considers the smaller value for f, lies between those found in Ref. [17], $\dot{I} <$ ln(1+4π).E/h, expressing the Shannon capacity of a quantum channel; and in Ref. [18], $\dot{I} <$ 2π².E/h.ln2, from the upper bound to the entropy/energy ratio for finite systems, being a factor 2/π apart from the last one.

The preceding result expressing a linear dependence of dI/dt with E corresponds to the case in which all the signal energy (to be precise, the deliverable energy, or the excess energy over the ground state) is delivered in one $\overline{\delta q}$ for in this case is E identifiable with $\overline{\delta H^e}$ in Eq. (66b). For signals lasting longer times it is convenient to rewrite Eq. (66) in terms of the rate at which the signal delivers its energy. Using Eq. (15) to replace $\overline{\delta H^e}$ in Eq. (66), the expression $\tau = f/k\theta$ instead of $1/\delta q \cdot dq/dt$, and multiplying both sides times $\tau$, one gets

$$\overline{\delta I} = \frac{-\overline{\delta H^e}}{k\theta \cdot \ln 2} = \frac{-f}{(k\theta)^2 \cdot \ln 2} \frac{dE}{dt} \quad . \quad (68)$$

The first equality expresses the energetic cost per bit $k\theta \cdot \ln 2$, as established by Brillouin [15]. Considering now that $-\overline{\delta H^e} \leq k\theta/2$, as this is the maximum energy contained in one $\delta q$, or, equivalently, deliverable in one $\tau$, two inequalities follow:

$$-\frac{dE}{dt} \leq \frac{(k\theta)^2}{2f} \quad , \quad (69a)$$

and $\quad \overline{\delta I} \leq \frac{1}{2\ln 2} bits \quad . \quad (69b)$

Eq. (69a) defines an upper bound to the rate at which the energy is delivered by a continuous signal per degree of freedom, and can be, therefore, compared with results concerning the capacity of what is called a single quantum channel. In terms of this rate, Eq. (67) rewrites

$$\dot{I} < \frac{1}{\ln 2}\left(-\frac{1}{2f}\frac{dE}{dt}\right)^{\frac{1}{2}} < \frac{1}{\ln 2}\left(-\frac{1}{\hbar}\frac{dE}{dt}\right)^{\frac{1}{2}},$$
(70)

which is, out of a factor $(\pi/3)^{1/2}$, the accepted expression for a continuous signal [36], consistent with results of quantum thermal conductance measurements [37]. The comparability of the upper bounds for the information rate obtained through so different methods relies on the independence of the calculated bounds from material properties, particle statistics, and E-k relationship as demonstrated in Refs. [36,38].

 First inequality in Eqs. (67) and (70) indicate that not only the energy, but also the actual f-value, determines the upper bound for the rate of information flow which may be, therefore, significantly lower than the ultimate bounds given in the right hand member or in Refs [17,18,36].

## III. SUMMARY AND DISCUSSION

1) We started from a mathematical property of the Lagrangian function $L(q, \dot{q}, t)$: Its first order variation is zero when the actual trajectory is varied in the direction $(\delta q, \delta \dot{q})$ satisfying $p \cdot \delta q = \text{const}$. This property serves to determine the actual trajectory for

-A) first order variation equals zero $\Longleftrightarrow$

$$\frac{\partial L}{\partial q}.\delta q + \frac{\partial L}{\partial \dot{q}}.\delta \dot{q} = 0$$

-B) $p \cdot \delta q = f = \text{const}$
$\Longleftrightarrow$
$$\dot{p}.\delta q + p.(\dot{\delta q}) = 0 \quad .$$

Being $\dot{\delta q} = (\dot{\delta q})$ and $p \equiv \partial L/\partial \dot{q}$, A and B yield $\frac{d}{dt}(\frac{\partial L}{\partial \dot{q}}) - \frac{\partial L}{\partial q} = 0$, the Lagrange equation for the classical trajectory.

2) We searched for a physical interpretation of the above. We proposed that first order variation equal zero, which indicates that the whole $\delta L$ (not only its first order variation) is close to zero for small deviations, means that trajectories separated by $\delta q$ from the classical "actual" trajectory are not distinguishable from each other, or that $\delta q$ measures the trajectory uncertainty if the constant f is small enough. This version, stated "The 'actual' classical trajectory has an associated uncertainty $\delta q$ given by CPDQ with f a small constant", satisfies A) and B) above because

(B) is expressed by the more restricted $p \cdot \delta q = f$, a small enough constant

(A) is implied in: "$\delta q$ is the trajectory uncertainty" $\Rightarrow \delta L \approx 0 \Rightarrow$ first order $\delta L = 0$.

3) Having introduced the spatial indeterminacy in the particle -- degree of freedom-- motion, or

the minimum uncertainty with which the rest of the world perceives this motion, and having shown the way it determines the classical trajectory, we shown in the rest of the article how a δq based description (uncertainty based description) becomes a common ground for the different branches of physics (CM, TD and QM) and provides, in addition, the criterion for distinguishing equilibrium from non-equilibrium processes.

The above can be summarized in the following expression

$$f \equiv p\delta q = \frac{\hbar}{2} e^{\frac{S}{k}} \qquad (71)$$

being p the classical q-conjugated momentum, and S the entropy, a measure of how much hidden from direct observation/interaction an enclosed particle is (S=0 for directly interacting degrees of freedom); and appropriate representations for δq as follows:

Classical Mechanics: S=0 (CPDQ) and smoothly varying δq ($\overline{\delta\delta q} << \delta q$). δq represented by a trajectory variation which zeroes the corresponding δL (indeterminacy), leads to Lagrange equations of motion (Section II.B)

Quantum Mechanics: S=0 and not necessarily smoothly varying δq: Not describable by a continuous succession of classical "states" (defined p and δq), or trajectory.. δq represented by the characteristic length of a complex probability amplitude ψ(q,t), which is, as a consequence, the representation of the state of motion. The representation is given by the lowest generalized Cramer-Rao bound --Eq. (54)--, which together with Eq. (71) lead to Schrödinger Equation. (Section II.K)

Thermodynamics (equilibrium): S not zero (enclosed particle). δq given by the enclosure dimensions (δq α $V^{1/3}$), leads, replaced in Eq. (71), directly to the the entropy of an ideal gas, and, indirectly, to the fundamental relationship between mechanical and thermodynamic quantities (heat theorem) after the identification θ = 1/k . p.dq/dt for the temperature (Section II.G). Non equilibrium thermodynamics (not fullfilling $\overline{\delta\delta q} << \delta q$ and $\overline{\delta p} << p$) is not addressed in this work.

4) Having, in all cases, interpreted δq as uncertainty, it becomes ineluctable to discuss CPDQ with reference to Heisenberg principle of uncertainty (HPU). We are faced to the following expressions: p.δq= $\hbar$/2 (CPDQ), and Δp.Δq≥ $\hbar$/2 (HPU).

We will first discuss the meaning of the quantities involved in each expression.

δq is, in CPDQ, a property of the trajectory. For each q(t) there is an associated δq(t), no measurement involved. Although we know the uncertainty associated to each position, we do not known, from the knowledge of the trajectory, where the particle is unless initial conditions (a measurement) are given. In HPU Δq is the uncertainty in position after a measurement was performed, and it is related with the uncertainty in momentum Δp after the same measurement.

CPDQ has a particular representation in semi-classical wave-mechanics in the de Broglie momentum-wavelength relationship (δq=$\lambda_B$/4π) as is obtained from Eq. (11), the found value $\hbar$/2 for the constant, and the definition of the de Broglie wavelength. But whereas the de Broglie wavelength is exclusively related to the undulatory description of the motion, δq (in CPDQ) is not necessarily associated with waves. Having established this relationship, one may go back to trajectories and say they have an uncertainty of the order of the de Broglie wavelength, a known fact. In order to emphasize the distinction, δq could be better called indeterminacy, as it denotes a property inherent in the thing; and Δq an uncertainty that is lack of knowledge about the thing.

Now, from the logical inequality Δp≤p, CPDQ and HPU, it follows Δq≥δq. I.e., at any given instant δq is always smaller than the uncertainty with which the coordinate q is "known" by whichever interacting system. In this sense δq may be considered the "size" of the particle and is, therefore, a lower limit to the size of the interval within which it makes sense to assert the particle is.

The lowest limit for this "size", and for the uncertainty Δq, occurs for the maximum p, limited in turn by relativistic considerations leading to $\delta q_{min}$= $\lambda_C$/4π, being $\lambda_C$=h/mc the Compton wavelength

5) Indeterminacy in the fundamental description enables the use of information theory framework. It has the advantage of making comparable whichever change in different systems in the same way as energetic terms make comparable changes involving work and heat, independently of the systems detail. In this framework a single quantity $\underline{\delta}I$, the elemental change (corresponding to $\overline{\delta}q$) of information serves to distinguish the domains of CM and QM processes, $\overline{\delta}I \equiv 0$, from TD ones; as well as to characterize equilibrium, or trajectory describable, processes $\overline{\delta}I << 1$, and to obtain from its limit value, $\overline{\delta}I < 1/2\ln 2$ bits, the upper bound for the information transfer rate.

**APPENDIX A. Classical limit with uncertainty. (Proof of consistency)**

Both representations of $\delta q$ -- indeterminacy in the classical trajectory, or (generalized) Fisher length-- should lead to the same expressions in case $\delta \bar{\delta} q << \delta q$, meaning mechanical quantities are well defined within $\delta q$. In this case, the knowledge of the potential energy $V(x)$ allows to write the quasi-Schrodinger Eq. (56)

$$-4C\psi''(x') + V(x)\psi'(x') = E.\psi'(x')$$

with general solution $\Psi(x,x') = A.\exp(ik(x).x') + B.\exp(-ik(x).x')$, where

$$k(x) = [(E - V(x))/4C]^{\frac{1}{2}}.$$

-Calculate the indeterminacy $\delta x$ using the lower limit in expression (54)

$$\delta x(x) = \left[\int 4\psi'^*\psi' dx\right]^{-\frac{1}{2}} = \frac{1}{2k(x)}$$

provided the integration limits are the same used in normalizing $\psi(x')$, and delimit an integer number of periods.

-Replace in the "uncertainty form" of Newton law, Eq. (18)

$$\dot{p} = -\frac{p.\dot{x}}{\delta x} \cdot \frac{\partial \delta x}{\partial x}$$

to get

$$\dot{p} = -\frac{\partial V}{\partial x} \quad .$$

**References and footnotes**

1. See for example H. Goldstein, *Classical Mechanics* (Addissson – Wesley, Massachusets, USA, 1950).
2. D. F. Styer et al., Am. J. of Physics **70**, 228 (2002).
3. See for example L. Landau and E. Lifschitz, *Physique Statistique* (Ed. MIR, Moscou, 1967).
4. J. W. Gibbs(1902), *Elementary Principles in Statistical Mechanics*, (Dover, N.Y. 1960).
5. Ehrenfest, P. and T. Ehrenfest-Afanassjewa (1912), *The Conceptual Foundations of the Statistical Approach in Mechanics* (Cornell University Press, New York 1959).
6. E. Madelung, Z. Phys. **40**, 322 (1926).
7. L. de Broglie, *The current interpretation of wave mechanics. A critical study* (Elsevier, Amsterdam, 1964).
8. D. Bohm, Phys. Rev. **85**, 166 (1952).
9. E. Nelson, Phys. Rev. **150**, 1079 (1966).
10. R. A. Fisher, *Statistical Methods and Scientific Inference* (Oliver and Boyd, London, 1959).
11. B. R. Frieden, Phys. Rev A **41**, 4265 (1990).
12. M. J. W. Hall, Phys. Rev. A **62**, 012107-1 (2000).
13. M. Reginatto, Phys. Rev. A **58**, 1775 (1998).
14. E. T. Jaynes, Phys.Rev. **106**, 620 (1957).
15. L. Brillouin, *La Science et la Théorie de l'Information* (Ed. Jacques Gabay, Sceaux, 1988, facsimile first edition Masson Ed. 1959).
16. B. Roy Frieden, B. H. Soffer, Phys. Rev. E, **52**, 2274 (1995).
17. H. J. Bremmerman, International Journal of Theoretical Physics, **21(3/4)**, 203 (1982).
18. J. D.Bekenstein, Phys. Rev. Lett. **46**, 623 (1981).
19. Being the trajectory variation $\delta q(t)$ a function of t, it implies also a variation of $\dot{q}(t)$. This is to note that when we refer to the special variation $\delta q$, its meaning is a variation in the specific direction $(\delta q, \delta \dot{q})$ on the plane $(q, \dot{q})$, the arguments of L.
20. This equality which will be discussed later holds only for the coordinates. In general $\overline{\delta}y \neq \delta y$ as occurs for example with L: $\delta L=0$ while $\overline{\delta}L$ is zero only for an inertial motion in absence of forces.

evident as stems from the formal relationship between Fisher information and Bohm's quantum potential, as pointed out in Refs. [12] and [13]. In particular Eq. (43) with V(x') instead of V(x) corresponds to Eq. (16) in Ref. [11]. Related ideas are also found in M.J.W. Hall and M. Reginatto, e-print quant-ph/0102069 (2002). Our approach is, however, conceptually different, proceeding straightforward from the same principle on which CM relies.